\documentclass{aa}
\usepackage[colorlinks,citecolor=blue,linkcolor=blue,urlcolor=blue]{hyperref}

\title{The sequence of Compton dominance in blazars based on data from WISE and Fermi/LAT}
\titlerunning{Compton dominance sequence of blazars}
\author{Krzysztof Nalewajko\thanks{\tt knalew@camk.edu.pl} and Maitrayee Gupta}
\authorrunning{Nalewajko \& Gupta}
\institute{Nicolaus Copernicus Astronomical Center, Polish Academy of Sciences, Bartycka 18, 00-716 Warsaw, Poland}

\abstract{
The two-component broad-band spectral energy distributions of blazars were suggested to form a sequence in which (1) the peak frequency of the low-energy (synchrotron) component $\nu_{\rm syn}$ is anticorrelated with the synchrotron luminosity $L_{\rm syn}$, (2) the luminosity ratio of the high-energy (inverse Compton) to synchrotron components $q = L_{\rm IC}/L_{\rm syn}$ (Compton dominance) increases with $L_{\rm syn}$ from the BL Lac objects (BL Lacs) to the flat spectrum radio quasars (FSRQs).  The Compton dominance parameter is an important probe of plasma magnetisation in the blazar zones within relativistic jets.  We investigate a sample of blazars detected by WISE in the mid-infrared (MIR) band and by Fermi/LAT in the GeV gamma-ray band, with the focus on the distribution of luminosities and photon indices.  Our findings are the following: (1) the MIR photon index $\Gamma_{\rm W12}$ is a useful probe of the blazar sequence, with the exception of low-luminosity BL Lacs that are most likely contaminated by their host galaxies ($L_{\rm W1} \sim 10^{44}\;{\rm erg\,s^{-1}}$ and $\Gamma_{\rm W12} < 1$); (2) $\Gamma_{\rm W12}$ is correlated with the gamma-ray photon index $\Gamma_{\rm 1-100\;GeV}$, with the MIR luminosity $L_{\rm W1}$, and with the Fermi/WISE Compton dominance $q_{\rm FW} = L_{\rm 1\;GeV}/L_{\rm W1}$; (3) a clean separation between FSRQs and BL Lacs can be seen in the parameter space of $\Gamma_{\rm W12}$ and $q_{\rm FW}$; (4) the observed distribution of MIR luminosity $L_{\rm W1}$ vs. Compton dominance $q_{\rm FW}$ for the entire sample of blazars can be modelled as a sequence of lepto-magnetic jet powers in the range $\log_{10} P_{\rm eB} \in [42:45]$ with the preference for sub-equipartition magnetic fields $P_{\rm B}/P_{\rm e} \in [0.05:1]$, assuming fixed bulk Lorentz factor $\Gamma_{\rm j} = 15$, fixed jet opening angle $\Gamma_{\rm j}\Theta_{\rm j} = 0.3$, fixed radiative efficiency of jet electrons $\epsilon_{\rm em} = 50\%$, and that external radiation luminosity scales like $L_{\rm ext} \propto P_{\rm eB}^{1.6}$ (parameter degeneracies are discussed).
}


\begin{document}

\maketitle

\section{Introduction}

Blazars belong to the most luminous cosmic sources in terms of apparent luminosity. Their spectral energy distributions are dominated by non-thermal emission forming two main components: the synchrotron component extending between the radio and UV bands (in the case of BL Lac objects extending to the X-ray band), and the inverse Compton component (in the leptonic scenarios) dominating in the gamma-ray band \citep[see][for recent review]{mad16}. In the basic leptonic models of blazar emission, both of these components are produced by the same population of ultra-relativistic electrons accelerated in the so-called blazar zone in a relativistic jet. The physics of energy dissipation and particle acceleration in relativistic jets remains elusive. Two particular mechanisms are discussed most frequently: relativistic shock waves and relativistic magnetic reconnection. It is thought that their efficiency can be determined mainly from the composition of jets, and particularly from their magnetisation $\sigma = B^2/(4 \pi w)$, where $w$ is the relativistic specific enthalpy. While shock waves are more efficient at low magnetizations ($\sigma < 1$), magnetic reconnection is more efficient at high magnetizations ($\sigma > 1$).
A very important argument was given by \cite{sir15}: both in the case of shocks and reconnection, the emitting regions are distinct from their background, and always closer to equipartition between the magnetic and kinetic energies ($\sigma \sim 1$).
Although observations of blazars have been argued to be roughly consistent with equipartition \citep{boe13,der14}, they cannot reveal the magnetisation of the background jet plasma. This also implies that the blazar emitting regions should not be characterised by high magnetisation values ($\sigma > 1$).

In luminous blazars, the dominant source of soft photons for Comptonization into the observed gamma-ray emission originates externally to the jet. Depending on the distance scale for the location of the emitting region along the jet, it could be either direct radiation from the accretion system \citep{der92}, broad emission lines \citep{sik94}, or thermal emission of the hot dust \citep{bla00}. In any case, the apparent gamma-ray luminosity is proportional to the energy density of the external radiation converted into the jet co-moving frame $L_\gamma \propto u_{\rm ext}'$. On the other hand, the apparent luminosity of the synchrotron component is proportional to the local magnetic energy density $L_{\rm syn} \propto u_{\rm B}'$. As has been discussed by \cite{nal14b}, the luminosity ratio of the Comptonization and synchrotron SED components $q = L_{\rm IC}/L_{\rm syn}$, referred to as the Compton dominance parameter, is an indirect probe of the jet magnetisation $q \simeq (\xi/0.005)(\Gamma/20)^2(\Gamma\theta_{\rm j})^2(L_{\rm d}/L_{\rm B})$, where $\xi$ is a parameter related to the covering factor of the medium reprocessing the accretion disk radiation of luminosity $L_{\rm d}$, and $L_{\rm B}$ is the magnetic jet power. Highly magnetised emitting regions in blazar jets are expected to produce radiation characterised by low Compton dominance $q \sim 1$ and high synchrotron luminosity $L_{\rm syn} > 10^{47}\;{\rm erg\,s^{-1}}$ \citep{jan15}. The eventual existence of such blazars would challenge the established theoretical expectation of rough equipartition in the blazar zone.

The spectral energy distributions of all types of blazars appear to be organised into so-called blazar sequence, the primary statement of which is an anticorrelation between the synchrotron SED peak frequency $\nu_{\rm syn}$ and the bolometric synchrotron luminosity $L_{\rm syn}$ \citep{fos98}. The second statement is that the Compton dominance $q$ increases with increasing $L_{\rm syn}$, i.e. that $L_{\rm IC}$ increases faster than $L_{\rm syn}$. Hence, at the low-luminosity end of the sequence we have so-called high-frequency peaked BL Lac objects (HBL) with $\nu_{\rm syn} \sim 10^{16}\;{\rm Hz}$, $L_{\rm syn} \sim 10^{44}\;{\rm erg\,s^{-1}}$ and $q \sim 1$; and at the high-luminosity end of the sequence we have flat-spectrum radio quasars (FSRQ) with $\nu_{\rm syn} \sim 10^{13}\;{\rm Hz}$, $L_{\rm syn} \sim 10^{47}\;{\rm erg\,s^{-1}}$ and $q \sim 10-100$. Since the synchrotron peak frequency for FSRQs is expected in the far-infrared band, it is very difficult to observe it directly, and it has been mostly determined by fitting simple SED templates to the radio/mm and NIR/optical data.

A sensitive all-sky survey in the mid-IR band has been performed by the WISE satellite in 2010 \citep{wise}. Most gamma-ray bright blazars can be associated with a WISE source. It has been demonstrated that blazars occupy a narrow region in a MIR colour-colour diagram \citep{mas11}, which has been very useful for identifying new blazar candidates. A combined WISE and Fermi/LAT sample of blazars was studied by \cite{dab12}, who analysed the observed distributions of infrared and gamma-ray colours, spectral indices, fluxes, as well as the Compton dominance parameter.

Here, we perform further analysis of a combined Fermi/LAT and WISE sample of blazars with known redshifts.
In Section \ref{sec_sample} we describe the selection of our sample.
In Section \ref{sec_obs} we present the distributions of physical parameters of blazar SEDs (luminosities, photon indices and Compton dominance) derived from infrared-only and combined infrared/gamma-ray data for our sample.
In Section \ref{sec_model} we present a very simple one-zone leptonic model of emission from relativistic jets, the predictions of which are compared directly with the observed distribution of synchrotron luminosity and Compton dominance.
We conclude with a brief discussion in Section \ref{sec_disc}.

To calculate the apparent luminosities of blazars, we adopted a standard $\Lambda$ cold dark matter cosmology with $H_0 = 71\;{\rm km\,s^{-1}}$, $\Omega_m=0.3$, and $\Omega_\Lambda=0.7$.

\section{WISE/Fermi blazar sample}
\label{sec_sample}

This study is based on the 2LAC catalog of gamma-ray bright blazars detected by Fermi/LAT between 2008 August and 2010 August \citep{2lac} crossed with the AllWISE Source Catalog of infrared point sources\footnote{\url{http://wise2.ipac.caltech.edu/docs/release/allwise/}}.
We have deliberately not used a more recent 3LAC catalog based on 4 years of Fermi/LAT data (2008 August - 2012 August) \citep{3lac}, since the most sensitive cryogenic part of the WISE survey lasted from 2010 January to 2010 September, and hence the 2LAC results are more likely to be simultaneous with WISE observations.

From 2LAC, we took 310 flat-spectrum radio quasars (FSRQs) and 175 BL Lac objects (BL Lacs) with known spectroscopic redshifts and searched for their infrared counterparts in the WISE catalog using the matching radius of 2 arcsec. This resulted in a sample of 152 FSRQs and 97 BL Lacs.

WISE observed the Universe in four infrared bands with the following central wavelengths:
$3.4\;{\rm\mu m}$ (W1),
$4.6\;{\rm\mu m}$ (W2),
$11.6\;{\rm\mu m}$ (W3),
$22.1\;{\rm\mu m}$ (W4).
In order to convert the measured magnitudes to fluxes, we use the zero-point fluxes provided by \cite{wise}:
$F_{\rm W1} = 307\;{\rm Jy}$,
$F_{\rm W2} = 171\;{\rm Jy}$,
$F_{\rm W3} = 29\;{\rm Jy}$,
$F_{\rm W4} =  8.3\;{\rm Jy}$.

\section{Observational results}
\label{sec_obs}

\subsection{Infrared properties}

\begin{figure*}
\centering
\includegraphics[width=\textwidth]{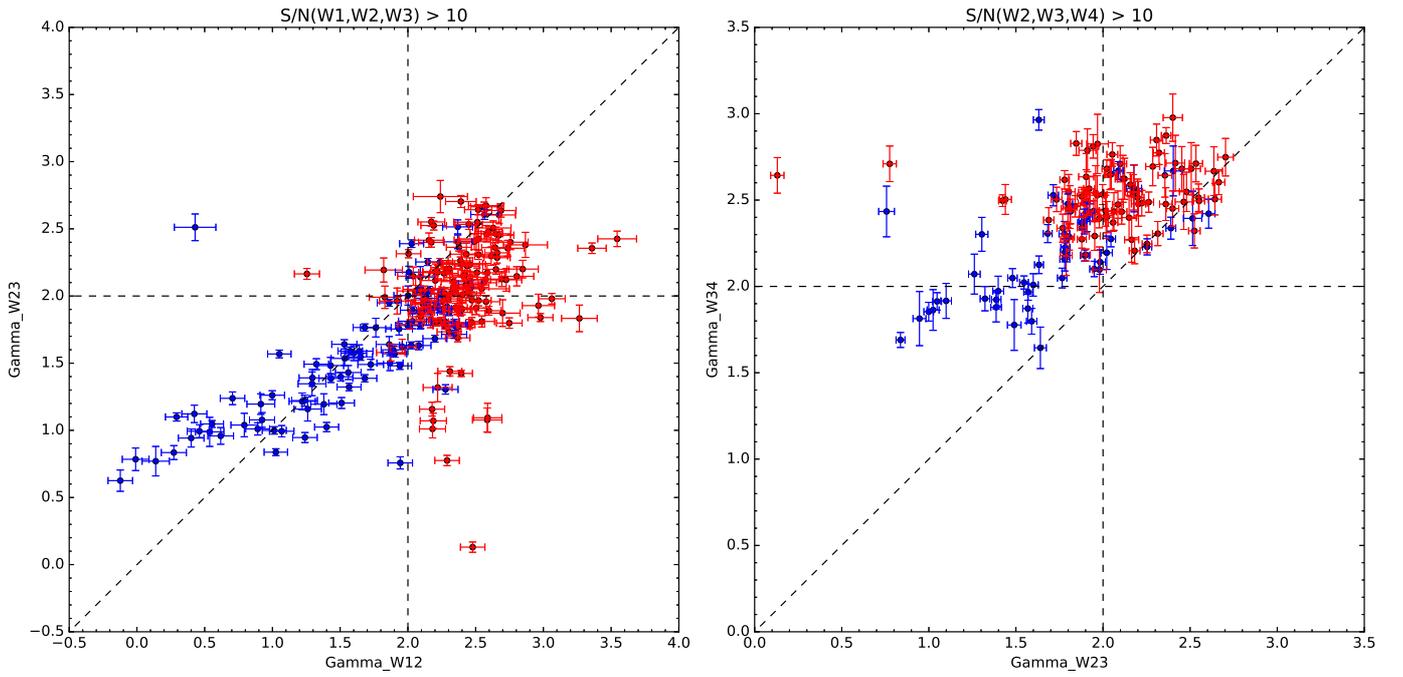}
\caption{Distribution of infrared photon index values of blazars determined from the WISE photometry. Left panel: $\Gamma_{\rm W12}$ vs. $\Gamma_{\rm W23}$. Right panel: $\Gamma_{\rm W23}$ vs. $\Gamma_{\rm W34}$. FSRQs (red) and BL Lacs (blue).}
\label{fig_wise_gamma_gamma}
\end{figure*}

Here we present results based solely on the WISE survey data. Our starting point is the WISE blazar strip established by \citep{mas11} using a colour-colour diagram $W_1-W_2$ vs. $W_2-W_3$. We translate this result into a photon index diagram $\Gamma_{12}$ vs. $\Gamma_{23}$ (Figure \ref{fig_wise_gamma_gamma}; left panel), using the standard definition of the photon index: $N(E) \propto E^{-\Gamma}$. In general, FSRQs are found to have soft infrared SEDs ($\Gamma > 2$) and BL Lacs have hard infrared SEDs ($\Gamma < 2$), which is in line with the blazar sequence \citep{fos98}, and also with its latest revision including the WISE data \citep{ghi17}. As for the spectral curvature, we adopt a convention that a positive curvature corresponds to the natural shape of the synchrotron SED: $\log N(E) = C - \alpha \log E - \beta \log^2E$. In the case of FSRQs, we find a preference for positive curvature with the mean value of $\Gamma_{12} - \Gamma_{23} = \beta \log(E_1/E_3) = 0.36$, hence $\beta \sim 0.68$. However, in the case of BL Lacs, we find a negative curvature for $\Gamma_{12} < 1.0$. Negative curvature suggests a superposition of two spectral components, most likely a contamination of the synchrotron component with radiation from the host galaxy.

If instead we look at the relation between between photon indices $\Gamma_{23}$ vs. $\Gamma_{34}$, for sources that are significantly detected ($S/N > 10$) in all $W_2,W_3,W_4$ bands, we find predominantly negative curvatures with $\Gamma_{34} > \Gamma_{23}$.
This is highly suspicious, and we suppose that the $W_4$ fluxes could be systematically overestimated.

Even though the W4 measurements are relatively poor, it is interesting to consider them as a possible probe of synchrotron self-absorption (SSA). The characteristic SSA frequency $\nu_{\rm SSA}$ is a sensitive probe of the radius $R$ of the emitting region, at the same time it depends only weakly on the jet Lorentz factor $\Gamma_{\rm j}$ and the synchrotron luminosity $L_{\rm syn}$. If the emitting region is located at relatively large distance scales ($r \sim 1-10\;{\rm pc}$), the SSA break can be located in the millimeter band \citep{sik08}.
However, in the case of very compact blazar emitting regions, e.g., located at distance scales within the broad-line region (BLR; $r < 0.1\;{\rm pc}$), the SSA break will be pushed to the MIR band \citep{hay12}.
If the SSA break could coincide with the W3 band, we would expect a very hard spectrum at longer wavelengths ($\Gamma_{34} < 1$) with normal spectrum at shorter wavelengths ($\Gamma_{23} \simeq 2$).
Looking at Figure \ref{fig_wise_gamma_gamma}, we do not have any indication for this.
On the other hand, several FSRQs are characterised by hard $\Gamma_{23} < 1.5$ and normal $\Gamma_{12} \simeq 2-2.5$, which could be a signature of an SSA break located near the W2 band.

\begin{figure*}
\centering
\includegraphics[width=0.5\textwidth]{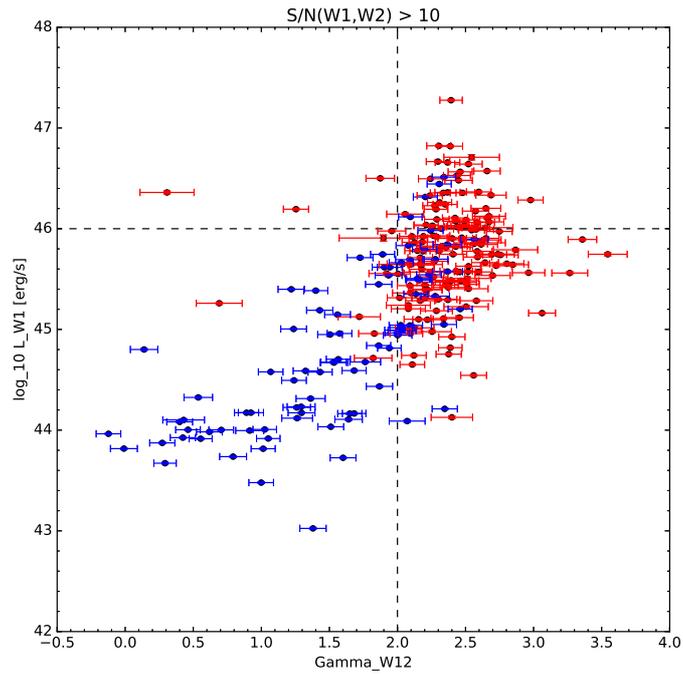}
\caption{Distribution of infrared photon index $\Gamma_{\rm W12}$ vs. infrared luminosity $L_{\rm W1}$ for blazars. FSRQs (red) and BL Lacs (blue).}
\label{fig_wise_gamma_lum}
\end{figure*}

In Figure \ref{fig_wise_gamma_lum}, we show the distribution of photon index $\Gamma_{12}$ plotted against the mid-infrared luminosity $L_1 = (\nu L_\nu)_{W1}$, calculated using spectroscopic redshifts provided in the 2LAC catalog. We first note that the mid-infrared luminosities of blazars reach values up to $L_{1,\rm max} \sim 10^{47}\;{\rm erg\,s^{-1}}$.
The luminosities of FSRQs extend in the range $45 < \log L_1 < 47$ and the luminosities of BL Lacs are in the range $44 < \log L_1 < 46.5$. The most luminous BL Lacs ($\log L_1 > 45.5$) overlap with the FSRQs and are characterised by relatively soft MIR spectra with $\Gamma_{12} \gtrsim 2$. Considering the entire sample of blazars, or BL Lacs alone, a trend of softer-when-brighter is clearly seen (Pearson $= 0.68$), consistent with the blazar sequence.
There are, however, significant outliers, e.g., a handful of luminous hard FSRQs with $\log L_1 > 46$ and $\Gamma_{12} < 2$.
These sources could be examples of luminous high-peaked blazars identified by \cite{pad12}.
The hardest BL Lacs, those with $\Gamma_{12} < 1$, appear to cluster around $\log L_1 \sim 44$. As we noted previously, these sources are also characterised with negative spectral curvature with $\Gamma_{23} > \Gamma_{12}$, and hence they could actually be dominated by contribution from the host galaxy.

We compared our distribution of $L_1$ luminosities with the synchrotron SED peaks $L_{\rm syn,peak,F13}$ calculated by modelling the broad-band SEDs with two-component log-polynomial models \citep{fin13}. With the spectral coverage of the entire infrared band being in general very poor, this method amounts to extrapolation of spectra measured in the radio/mm and NIR/optical/UV bands. We found that regardless of the order of magnitude for either luminosity estimate, the ratio $\log(L_1/L_{\rm syn,peak,F13})$ has approximately uniform distribution in the range $[-1.2:0]$. Hence, the extrapolated synchrotron peaks (their $\nu L_\nu$ rather than bolometric luminosities) exceed the observed MIR luminosities by up to one order of magnitude both for BL Lacs and FSRQs. The most likely reason for such discrepancy is that extrapolation of radio/mm or NIR/optical/UV spectra can be affected by additional spectral features, e.g., a synchrotron self-absorption break in the (sub)-mm band or the accretion disk continuum, especially at higher redshifts ($z \gtrsim 2$).

\subsection{Infrared vs. gamma-ray properties}

\begin{figure*}
\centering
\includegraphics[width=0.5\textwidth]{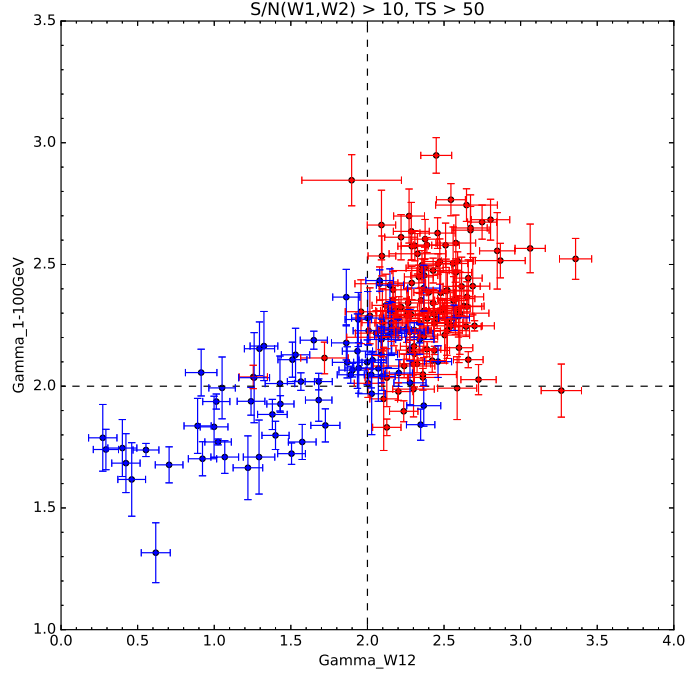}
\caption{Distribution of infrared photon index $\Gamma_{\rm W12}$ vs. gamma-ray photon index $\Gamma_{\rm 1-100GeV}$ for blazars. FSRQs (red) and BL Lacs (blue).}
\label{fig_wise_gammaIR_gammaLAT}
\end{figure*}

In Figure \ref{fig_wise_gammaIR_gammaLAT}, we show the distribution of gamma-ray photon index $\Gamma_{\rm 1-100\;GeV}$ plotted against the MIR photon index $\Gamma_{\rm W12}$ \citep[cf. Figure 6 in][]{dab12}. There is a clear correlation between the two indices (Pearson $= 0.69$), although the range of $\Gamma_{\rm 1-100\;GeV} \in [1.5:3]$ is more narrow. Most FSRQs are characterised both by soft MIR SED ($\Gamma_{\rm W12} > 2$) and soft gamma-ray SED ($\Gamma_{\rm 1-100\;GeV} > 2$). Most BL Lacs, excluding those that partially overlap with the FSRQs, are characterised by hard MIR SED ($\Gamma_{\rm W12} < 2$) and flat or moderately hard gamma-ray SED ($\Gamma_{\rm 1-100\;GeV} \lesssim 2$). Those BL Lacs with the hardest MIR SED ($\Gamma_{\rm W12} < 1$), that are suspected to be contaminated by their host galaxies, are characterised by $\Gamma_{\rm 1-100\;GeV} \simeq 1.7$.

In Figure \ref{fig_Lsyn_cd}, we show the distribution of the Compton dominance $q$ plotted against the MIR luminosity $L_{\rm W1}$. Compton dominance is calculated as $q_{\rm FW} = L_{\rm 1GeV}/L_{\rm W1}$, where $L_{\rm 1GeV}$ is the $\nu L_\nu$ luminosity determined from the power-law fit over photon energy range $1-100\;{\rm GeV}$ as provided in 2LAC. We find that most FSRQs are characterised by $1 < q_{\rm FW} < 10$, with only a few reaching values $q_{\rm FW} > 30$. On the other hand, most BL Lacs are characterised by $q_{\rm FW} < 1$, with the observational lower limit of $q_{\rm FW} > 0.1$. There is a substantial overlap of FSRQs and BL Lacs, with some FSRQs having $q_{\rm FW} \sim 0.3$ and some BL Lacs having $q_{\rm FW} \sim 5$. Considering the most synchrotron-luminous blazars with $L_{\rm W1} > 10^{46}\;{\rm erg/s}$, they span more than 2 orders of magnitude in Compton dominance with $0.3 < q_{\rm FW} < 50$. Sources with $L_{\rm W1} > 10^{46}\;{\rm erg/s}$ and $q_{\rm FW} < 1$ are potentially most interesting as sites of efficient dissipation in highly magnetised jet regions \citep{jan15}. In the next section, we will discuss a simple model of blazar emission from a relativistic jet that allows to calculate the luminosities of the synchrotron and inverse Compton components of the spectral energy distribution.

Figure \ref{fig_wise_index_q} shows the distribution of Compton dominance $q_{\rm FW}$ plotted against the MIR photon index $\Gamma_{\rm W12}$ or the gamma-ray photon index $\Gamma_{\rm 1-100\;GeV}$ \citep[cf. Figure 9 in][]{dab12}.
The separation between FSRQs and BL Lacs appear to be cleaner in the $\Gamma_{\rm W12}$ vs. $q_{\rm FW}$ space: most FSRQs are grouped in the region where $\Gamma_{\rm W12} \in [2:3)$ and $\log_{10} q_{\rm FW} \in [0:1.5)$, while most BL Lacs are found in the region where $\Gamma_{\rm W12} \in (0:2]$ and $\log_{10} q_{\rm FW} \in (-1:0]$.
A correlation between $\Gamma_{\rm 1-100\;GeV}$ and $q_{\rm FW}$ is rather weak (Pearson $= 0.26$).

\section{Model of jet energetics}
\label{sec_model}

We consider a simple one-zone leptonic model of blazar emission that can most naturally explain the distribution of apparent luminosities and Compton dominance along the blazar sequence. While this has been discussed in several previous studies \citep{mey11,fin13}, here we are guided by the following questions: (1) could blazar sequence be governed by the jet power, (2) what constraints can be put on the jet composition, i.e., the relation between magnetic and leptonic jet powers, (3) what is the role of the bulk Lorentz factor along the sequence?

We consider a jet of conical geometry with the bulk velocity $\beta_{\rm j} = v_{\rm j}/c$, the corresponding bulk Lorentz factor $\Gamma_{\rm j} = (1-\beta_{\rm j}^2)^{-1/2}$, and the opening angle $\theta_{\rm j}$.
The emitting region is fixed in the external frame and is approximated as a cylindrical shell with radius $R$ and length $\Delta r_{\rm em}$.
The shell radius is equal to the jet radius $R = r_{\rm diss}\theta_{\rm j}$ at the distance scale $r_{\rm diss}$, and the length $\Delta r_{\rm em} = \Gamma_{\rm j} c\tau_{\rm cool}'$ is related to the co-moving cooling time scale $\tau_{\rm cool}'$.
We consider a population of relativistic electrons (or positrons) of co-moving number density $n_{\rm e}'$ and mean Lorentz factor $\gamma_{\rm e}'$, and their sub-population of \emph{emitting electrons} of density $n_{\rm em}' < n_{\rm e}'$ and typical Lorentz factor $\gamma_{\rm em}'$ such that $\gamma_{\rm e}' < \gamma_{\rm em}' < (n_{\rm e}'/n_{\rm em}')\gamma_{\rm e}'$.
Every emitting electron produces radiation isotropic in the co-moving frame with emission power $P_{\rm em,1}'$ such that its radiative cooling time scale is $\tau_{\rm cool}' = \gamma_{\rm em}m_{\rm e}c^2/P_{\rm em,1}'$.
Transforming the total emission power $P_{\rm em}' = n_{\rm em}'V_{\rm em}'P_{\rm em,1}'$ into the external frame, we note that (1) the effective length of the emitting region in the co-moving frame is $\Delta r_{\rm em}' = \Delta r_{\rm em}/\Gamma_{\rm j} = c\tau_{\rm cool}'$, (2) hence the volume transformation is $V_{\rm em}' = \pi R^2\Delta r_{\rm em}' = V_{\rm em}/\Gamma_{\rm j}$, (3) the number density of electrons transforms like $n_{\rm e(em)} = \Gamma_{\rm j}n_{\rm e(em)}'$, and hence (4) the effective number of emitting electrons is $N_{\rm em} = n_{\rm em}V_{\rm em} = \Gamma_{\rm j}^2N_{\rm em}'$, (5) the power emitted by each electron is invariant $P_{\rm em,1} = P_{\rm em,1}'$, thus (6) the result is $P_{\rm em} = \Gamma_{\rm j}^2P_{\rm em}'$.
This can be compared with the total electron energy flux $P_{\rm e} = \pi R^2\Gamma_{\rm j}^2\gamma_{\rm e}'n_{\rm e}'m_{\rm e}c^3$, and we find that $P_{\rm em}/P_{\rm e} = (\gamma_{\rm em}/\gamma_{\rm e})(n_{\rm em}'/n_{\rm e}') \equiv \epsilon_{\rm em}$, which we call the radiation efficiency of jet electrons.

The emitted radiation in the external frame is strongly anisotropic due to (1) the relativistic Doppler effect on photon energy $E_\gamma = \mathcal{D}E_\gamma'$, and due to (2) the relativistic aberration affecting the solid angle ${\rm d}\Omega = {\rm d}\Omega'/\mathcal{D}^2$ or the cosine of the viewing angle $\mu_{\rm obs} = \cos\theta_{\rm obs}$ so that ${\rm d}\mu_{\rm obs} = {\rm d}\mu_{\rm obs}'/\mathcal{D}^2$, where $\mathcal{D} = [\Gamma_{\rm j}(1-\beta_{\rm j}\mu_{\rm obs})]^{-1} = \Gamma_{\rm j}(1+\beta_{\rm j}\mu_{\rm obs}')$ is the relativistic Doppler factor.
The apparent luminosity is $L_{\rm obs}(\mu_{\rm obs}) = (\mathcal{D}^3/\Gamma_{\rm j})P_{\rm em} = \mathcal{D}^3\Gamma_{\rm j}P_{\rm em}'$, so that $\left<L_{\rm obs}\right>_\Omega = P_{\rm em}$ \citep{sik97,jes08}.

We now consider three main leptonic radiative mechanisms that are relevant for blazars: synchrotron, synchrotron self-Compton (SSC) and external radiation Comptonization (ERC). For either of them, the emission power of a single electron is $P_{\rm em,1}' = (4/3)\sigma_{\rm T}c\gamma_{\rm em}^2u_0'$, where $u_0'$ stands for magnetic energy density $u_{\rm B}' = B'^2/(8\pi)$ in the case of synchrotron, energy density of synchrotron radiation $u_{\rm syn}'$ in the case of SSC, and external radiation density $u_{\rm ext}'$ in the case of ERC.
We can relate the magnetic energy density to the magnetic jet power $P_{\rm B} = \pi R^2\Gamma_{\rm j}^2u_{\rm B}'c$, and the external radiation energy density can be expressed as $u_{\rm ext}' \simeq \mathcal{D}^2L_{\rm ext}/(4\pi cr_{\rm ext}^2)$ \citep{der95}, where $L_{\rm ext}$ is the luminosity of external radiation sources or characteristic radius $r_{\rm ext} \sim r_{\rm diss}$.
The co-moving energy density of synchrotron radiation is approximately $u_{\rm syn}' \simeq g_{\rm ssc}P_{\rm syn}'/(4\pi cR^2)$, where $P_{\rm syn}' = P_{\rm syn,1}'n_{\rm em}'V_{\rm em}'$ and $g_{\rm ssc}$ is a geometric factor.
The value of $g_{\rm ssc}$ depends on the shape (mainly the aspect ratio) of the synchrotron emitting region in the co-moving frame, light travel effects and anisotropy of IC scattering, it is therefore difficult to calculate and quite uncertain.
The effective co-moving radiative cooling time scale is now $\tau_{\rm cool}' = 3m_{\rm e}c/(4\sigma_{\rm T}\gamma_{\rm em}u_{\rm tot}')$, where $u_{\rm tot}' = u_{\rm B}' + u_{\rm syn}' + u_{\rm ext}'$, and the respective luminosities are $L_{\rm syn} = (u_{\rm B}'/u_{\rm tot}')L_{\rm obs}$, $L_{\rm SSC} = (u_{\rm syn}'/u_{\rm tot}')L_{\rm obs}$, and $L_{\rm ERC} = (u_{\rm ext}'/u_{\rm tot}')L_{\rm obs}$.

We now consider two components of the Compton dominance parameter:
\begin{eqnarray}
q_{\rm SSC} &=& \frac{L_{\rm SSC}}{L_{\rm syn}} = \frac{u_{\rm syn}'}{u_{\rm B}'} = \frac{g_{\rm ssc}\epsilon_{\rm em}}{4}\frac{L_{\rm syn}}{L_{\rm obs}}\frac{P_{\rm e}}{P_{\rm B}}\,,
\\
q_{\rm ERC} &=& \frac{L_{\rm ERC}}{L_{\rm syn}} = \frac{u_{\rm ext}'}{u_{\rm B}'} = \left(\frac{\mathcal{D}\Gamma_{\rm j}\theta_{\rm j}}{2}\right)^2\frac{L_{\rm ext}}{P_{\rm B}}\,.
\end{eqnarray}
The observational trend for Compton dominance can be approximated as flat $q_{\rm SSC}$ and at least linearly increasing $q_{\rm ERC}$ over 3 orders of magnitude in $L_{\rm syn}$.
Let us now consider how these functions should scale with the \emph{lepto-magnetic jet power} $P_{\rm eB} = P_{\rm e} + P_{\rm B}$ (this can be generalised by including the contribution from protons) and bulk Lorentz factor $\Gamma_{\rm j}$.
As is typical, we assume that $P_{\rm e} \propto P_{\rm B} \propto P_{\rm eB}$, $\mathcal{D} \propto \Gamma_{\rm j}$, $\theta_{\rm j} \propto 1/\Gamma_{\rm j}$, $R \propto r_{\rm diss}/\Gamma_{\rm j}$.
We also assume that $\epsilon_{\rm em}$ is comparable for FSRQs and BL Lacs.

When the total emission is dominated by synchrotron (and SSC), as in the case of BL Lacs, $q_{\rm SSC} \propto P_{\rm e}/P_{\rm B}$.
On the other hand, when the total emission is dominated by ERC, as in the case of FSRQs, we find that $q_{\rm SSC} \propto P_{\rm e}/(q_{\rm ERC}P_{\rm B})$.
Hence, the energy balance between electrons and magnetic fields, i.e., the equipartition condition, is key in determining the value of $q_{\rm SSC}$.

The value of $q_{\rm ERC}$ depends primarily on the relation between external radiation luminosity $L_{\rm ext}$ and the magnetic jet power $P_{\rm B}$.
It is typically assumed that external radiation fields result from reprocessing of accretion flow luminosity $L_{\rm acc}$ by external medium (whether the broad line region or the dusty torus) of covering factor $\xi_{\rm ext} \sim 0.1$, such that $L_{\rm ext} = \xi_{\rm ext}L_{\rm acc}$.
The accretion luminosity is related to the accretion power $P_{\rm acc} \equiv \dot{M}_{\rm acc}c^2$ via radiative efficiency $\epsilon_{\rm acc}$, such that $L_{\rm acc} = \epsilon_{\rm acc}P_{\rm acc}$.
And the accretion power can be related to the lepto-magnetic jet power by parameter called jet production efficiency $\eta_{\rm eB} = P_{\rm eB}/P_{\rm acc}$.
While we cannot directly estimate the value of $P_{\rm acc}$, we can constrain the combination of parameters $\xi_{\rm ext}\epsilon_{\rm acc}/\eta_{\rm eB}$ by postulating a systematic relation between luminosity of external radiation and the total jet power $L_{\rm ext} = A(P_{\rm eB}/P_{\rm Edd})^\kappa$, where $P_{\rm Edd} = 1.5\times 10^{47}\;{\rm erg\,s^{-1}}$ is the Eddington luminosity for supermassive black hole of mass $M_{\rm bh} = 10^9 M_\odot$.
The value of $\kappa$ can be in general constrained to be within the range $1 < \kappa < 2$.
If $\kappa = 1$, we would predict that $q_{\rm ERC}$ is independent of the jet power.
On the other hand, if $\kappa = 2$, the value of $q_{\rm ERC}$ would become independent of $L_{\rm syn}$.

Using this scheme, we seek a reference model of blazar sequence, in which most parameters are tied to the lepto-magnetic jet power $P_{\rm eB}$.
For all models presented here, we consider the range of $42 < \log_{10} P_{\rm eB} < 45$.
For the reference model, we adopt the following key parameters: jet Lorentz factor $\Gamma_{\rm j} = 15$, jet opening angle $\theta_{\rm j} = 0.3/\Gamma_{\rm j}$, radiative efficiency $\epsilon_{\rm em} = 0.5$.

First, in order to match observed distribution of Compton dominance values for BL Lacs, i.e., the $q_{\rm SSC}$, we will adjust the jet magnetisation described by the $P_{\rm B}/P_{\rm e}$ ratio.
We have seen that $P_{\rm B}/P_{\rm e}$ affects $q_{\rm SSC}$, whether cooling is dominated by synchrotron or ERC.
Moreover, if we limit ourselves to the case where $P_{\rm B} < P_{\rm e}$, since $L_{\rm ext}$ is scaled with $P_{\rm eB} = P_{\rm B} + P_{\rm e} \sim P_{\rm e}$, the value of $q_{\rm ERC}$ will also be affected.
We find that jet magnetisation in the range $0.07 < P_{\rm B}/P_{\rm e} < 0.7$ can explain the observed values of $q_{\rm SSC}$ most naturally. Hence, we adopt a reference value of $P_{\rm B}/P_{\rm e} = 0.2$.

\begin{figure*}
\centering
\includegraphics[width=0.32\textwidth]{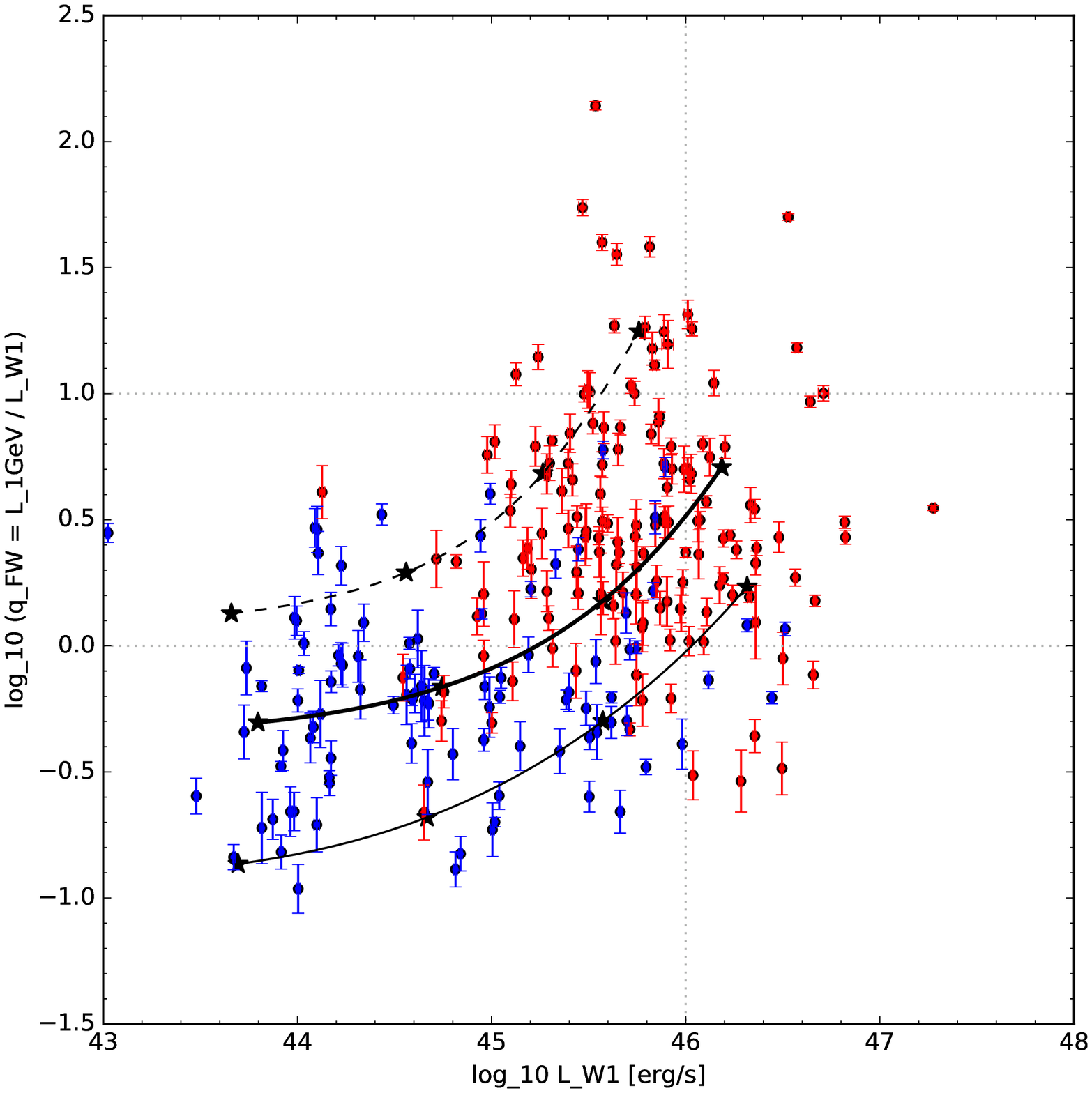}
\includegraphics[width=0.32\textwidth]{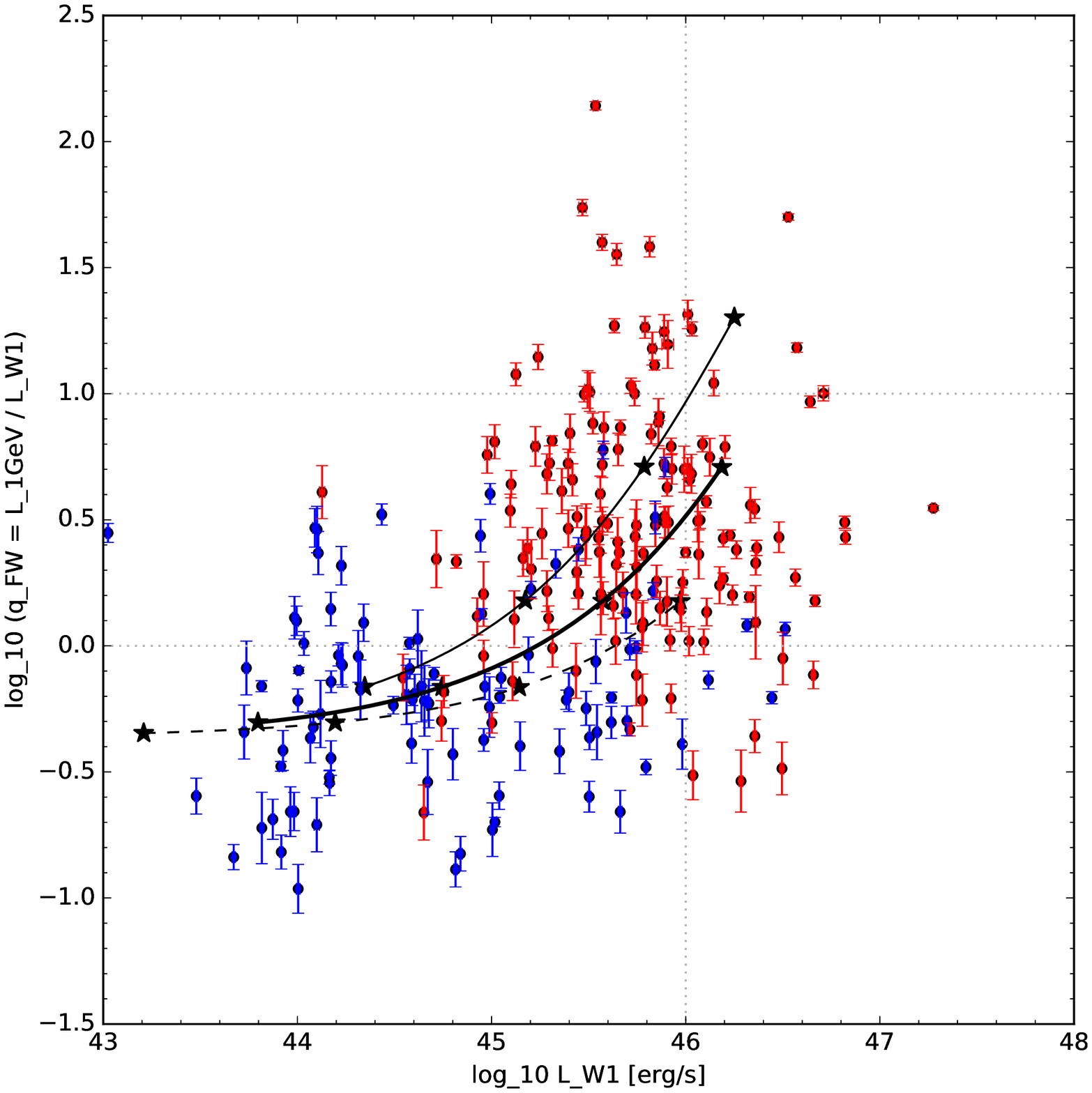}
\includegraphics[width=0.32\textwidth]{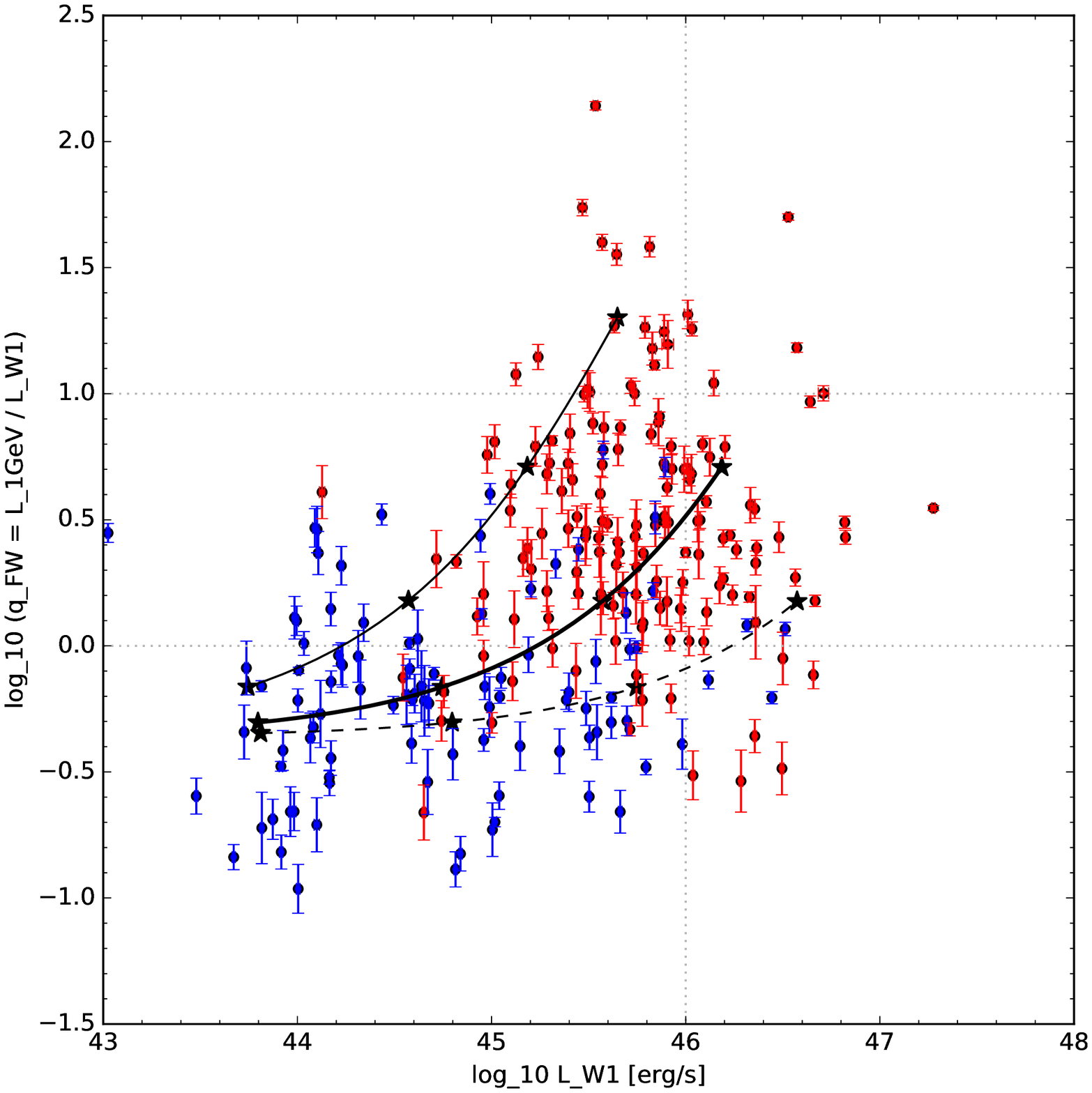}
\caption{Distribution of Fermi+WISE blazars in the parameter space of synchrotron luminosity $L_{\rm syn}$ and Compton dominance $q_{\rm FW} = L_\gamma/L_{\rm syn}$, where $L_\gamma = L_{\rm SSC} + L_{\rm ERC}$ (FSRQs - red, BL Lacs - blue).
Thick solid lines show our reference model of the blazar sequence for $\Gamma_{\rm j} = 15$, $P_{\rm B}/P_{\rm e} = 0.2$, $\Gamma_{\rm j}\theta_{\rm j} = 0.3$, and $\epsilon_{\rm em} = 0.5$.
Left panel: dependence of the model on the jet magnetisation $P_{\rm B}/P_{\rm e} = 0.05$ (dashed), 1 (thin solid).
Middle panel: dependence of the model on the jet Lorentz factor $\Gamma_{\rm j} = 7.5$ (dashed), 30 (thin solid).
Right panel: dependence of the model on the jet collimation factor $\Gamma_{\rm j}\theta_{\rm j} = 0.15$ (dashed), 0.6 (thin solid).
The black stars along each track indicate the lepto-magnetic jet power values $\log_{10}P_{\rm eB} = 42,43,44,45$.}
\label{fig_Lsyn_cd}
\end{figure*}

\begin{figure*}
\centering
\includegraphics[width=\textwidth]{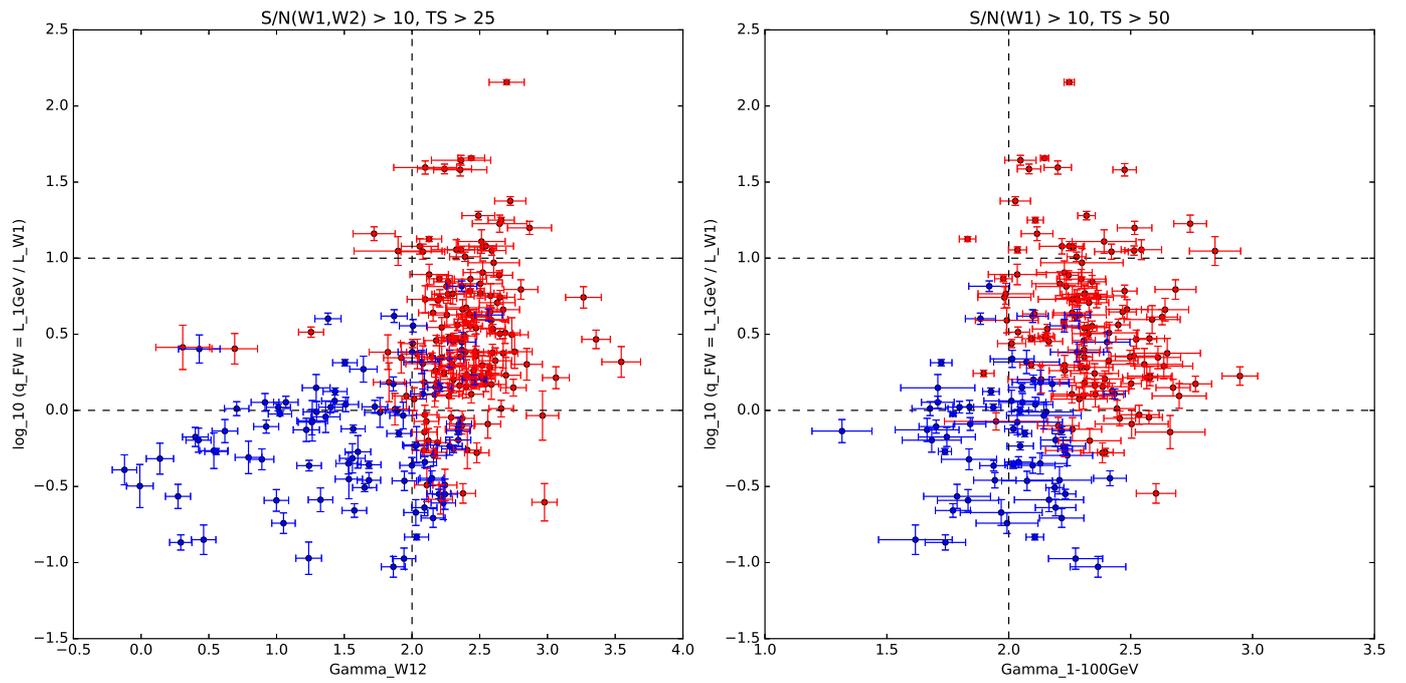}
\caption{Distribution of Compton dominance $q_{\rm FW} = L_{\rm 1-100GeV}/L_{\rm W1}$ vs. infrared photon index $\Gamma_{\rm W12}$ (left panel) and gamma-ray photon index $\Gamma_{\rm 1-100GeV}$ (right panel) for blazars. FSRQs (red) and BL Lacs (blue).}
\label{fig_wise_index_q}
\end{figure*}

Second, in order to match the observed distribution of $q_{\rm ERC}$, we normalise the luminosity of external radiation as $L_{\rm ext} \sim 5\times 10^{47}(P_{\rm eB}/P_{\rm Edd})^{1.6}\;{\rm erg\,s^{-1}}$, hence we adopt $A = 3.3P_{\rm Edd}$ and $\kappa = 1.6$.
Our reference model is now defined completely, and it is shown on every panel of Figure \ref{fig_Lsyn_cd} with thick solid lines.

Next, we consider the effects of jet Lorentz factor $\Gamma_{\rm j}$ and the jet collimation parameter $\Gamma_{\rm j}\theta_{\rm j}$.
In the middle panel of Figure \ref{fig_Lsyn_cd}, we show models obtained from the reference models by setting $\Gamma_{\rm j} = 7.5, 30$, keeping fixed $\Gamma_{\rm j}\theta_{\rm j} = 0.3$.
On the other hand, in the right panel of Figure \ref{fig_Lsyn_cd}, we show models corresponding to $\Gamma_{\rm j}\theta_{\rm j} = 0.15,0.6$, keeping fixed $\Gamma_{\rm j} = 15$.
We note that in the case of FSRQs, where the total radiation output is dominated by ERC, 
the synchrotron luminosity scales like $L_{\rm syn} \propto (\mathcal{D}/\Gamma_{\rm j})(\Gamma_{\rm j}\theta_{\rm j})^{-2}$, and Compton dominance scales like $q_{\rm ERC} \propto (\mathcal{D}\Gamma_{\rm j}\theta_{\rm j})^2$.
Hence, the effect of jet collimation parameter is stronger, as it affects both $L_{\rm syn}$ and $q_{\rm ERC}$.
Adopting a larger value of $\Gamma_{\rm j}\theta_{\rm j}$ allows to achieve higher Compton dominance for the same value of $P_{\rm B}/P_{\rm e}$, or to achieve the same value of $q_{\rm ERC}$ for higher value of $P_{\rm B}/P_{\rm e}$.

\section{Discussion}
\label{sec_disc}

We considered a simple parametrisation of blazar emission models in order to understand the observed distribution of Compton dominance along the blazar sequence. In our model, most parameters are tied to the lepto-magnetic jet power $P_{\rm eB}$, the value of which extends over 3 orders of magnitude to reproduce the observed range of synchrotron luminosities $L_{\rm syn}$. Theoretical Compton dominance has 2 components: $q_{\rm SSC} = L_{\rm SSC}/L_{\rm syn}$ and $q_{\rm ERC} = L_{\rm ERC}/L_{\rm syn}$, and under most plausible circumstances they both scale with the equipartition parameter $P_{\rm B}/P_{\rm e}$. In our reference model with $\Gamma_{\rm j} = 15$, $\Gamma_{\rm j}\theta_{\rm j} = 0.3$, $\epsilon_{\rm em} = 0.5$, and $L_{\rm ext}/P_{\rm Edd} \sim 3.3(P_{\rm eB}/P_{\rm Edd})^{1.6}$, the observed distribution of $L_\gamma/L_{\rm syn}$ can be reproduced for $0.05 < P_{\rm B}/P_{\rm e} < 1$, which means a jet dominated by matter (even without counting protons).
A very similar conclusion for the case of BL Lacs was obtained recently by \cite{tav16}. 
Our reference model also implies typical values for the lepto-magnetic jet power: $\log_{10} P_{\rm eB} \sim 42-43.5$ in the case of BL Lacs, and $\log_{10} P_{\rm eB} \sim 43.5-45$ in the case of FSRQs.
The highest gamma-ray luminosities achieved by the FSRQs, $L_{\rm 1\;GeV,max} \sim 10^{48}\;{\rm erg\,s^{-1}}$ (long-term average), imply the existence of upper limit on the lepto-magnetic power of radiatively efficient relativistic AGN jets $P_{\rm eB,max} \sim 2\times 10^{45}\;{\rm erg\,s^{-1}} \simeq P_{\rm Edd}/75$, leaving plenty of room for protons.
On the other hand, the highest observed synchrotron luminosities are systematically lower, with $L_{\rm W1,max} \sim 10^{47}\;{\rm erg\,s^{-1}} \sim L_{\rm 1\;GeV,max}/10$, much lower than values predicted for powerful jets under magnetisation $\sigma \sim 1$ \citep{jan15}.

The existence of upper limit on the jet power allows us to interpret the scaling of external radiation luminosity with the total jet power $L_{\rm ext}/P_{\rm Edd} \sim 3.3(P_{\rm eB}/P_{\rm Edd})^{1.6}$. Assuming that $\xi_{\rm ext},\eta_{\rm eB} = {\rm const}$, and that $\epsilon_{\rm acc} = \epsilon_{\rm acc,0}(P_{\rm acc}/P_{\rm acc,max})^{0.6}$, where $P_{\rm acc,max} = P_{\rm eB,max}/\eta_{\rm eB}$, we find that $L_{\rm acc}/P_{\rm Edd} \simeq 0.03\xi_{\rm ext,-1}^{-1}(P_{\rm acc}/P_{\rm acc,max})^{1.6}$ and that $\xi_{\rm ext}\epsilon_{\rm acc,0}/\eta_{\rm eB} \simeq 0.25$. This shows that for the most powerful blazars with $P_{\rm eB} \sim P_{\rm eB,max}$, the requirements for radiative efficiency of accretion and the covering factor of radiation reprocessing medium are very tight.

Degeneracy of parameters is widely recognised as the main obstacle in constraining the physical parameters of relativistic blazar jets. The SSC branch of Compton dominance $q_{\rm SSC}$ can be affected by radiative efficiency of jet electrons $\epsilon_{\rm em}$, while the ERC branch depends sensitively on the jet Lorentz factor $\Gamma_{\rm j}$, the jet collimation parameter $\Gamma_{\rm j}\theta_{\rm j}$, and on the scaling of external radiation luminosity $L_{\rm ext}$. However, the relatively limited range of observed values of Compton dominance, factor $\sim 100$ in the case of $q_{\rm ERC}$ and factor $\sim 30$ in the case of $q_{\rm SSC}$, means that there should be strict limits on the combinations of parameters like $(\mathcal{D}\Gamma_{\rm j}\theta_{\rm j})^2(L_{\rm ext}/P_{\rm B})$. This also means that while equipartition can be realised for FSRQs \citep[e.g.,][]{der14} by increasing $\Gamma_{\rm j}$ or $\Gamma_{\rm j}\theta_{\rm j}$, it would be difficult to obtain for BL Lacs, as the adopted value of electron radiative efficiency $\epsilon_{\rm em} = 50\%$ is already high.
Nevertheless, it has been suggested that magnetizations in blazar jets could be very high, with $P_{\rm B} \gg P_{\rm e}$, if the overall radiative efficiency (including magnetic fields but not including protons) is as high as 95\% \citep{pot17}.
Also, the value of geometric factor $g_{\rm SSC}$ is quite uncertain. High values of $g_{\rm SSC}$ could be obtained if the synchrotron emitting region is relatively thin compared with the jet radius $R$. This is the usual assumption of numerical models of blazar emission \citep{mod03}, and hence such models are essential to constrain the value of $g_{\rm SSC}$.

Still, our very simplified model suggests that it is probably easier and more natural to explain the observed distribution of blazar SEDs as function of $P_{\rm eB},P_{\rm B}/P_{\rm e}$ for a single value of $\Gamma_{\rm j} \simeq 15$ (as suggested by \citealt{ghi14} for the case of FSRQs), rather than a function of $P_{\rm eB},\Gamma_{\rm j}$ for a single value of $P_{\rm B}/P_{\rm e} \simeq 0.2$.
In the former scenario, the transition between SSC and ERC domination is roughly at the same value of $P_{\rm eB}$, while in the latter scenario some BL Lacs are predicted to be contaminated by ERC at high values of $\Gamma_{\rm j}$.
On the other hand, interferometric radio observations of blazars indicate that FSRQs have systematically higher jet Lorentz factors than BL Lacs \citep[e.g.,][]{hov09}.

Regarding the location of the emitting region \citep[see][for more details in the case of FSRQs]{nal14a}, we assumed that the energy density of external radiation fields scales like $u_{\rm ext}' = \Gamma_{\rm j}^2L_{\rm ext}/(4\pi cr_{\rm diss}^2)$, a simple power-law dependence on distance scale $r_{\rm diss}$. In reality, we can expect several contributions to the external photon fields with certain characteristic distance scales, e.g., $r_{\rm BLR} \sim 0.1L_{\rm acc,46}^{1/2}\;{\rm pc}$ for broad emission lines or $r_{\rm HDR} \sim 3L_{\rm acc,46}^{1/2}\;{\rm pc}$ for the hot-dust region (torus). Also, effects of special relativity can modify the inverse-square law for radiation density. Nevertheless, the inverse-square law applies roughly over several orders of magnitude in $r_{\rm diss}$, since in quasars the external radiation fields originate mostly from reprocessing of direct radiation of the innermost accretion disk \citep{sik09}.

Of course, the observed distribution of Compton dominance values could be subject of various selection effects \citep{gio12,fin13}. On one hand, highly Compton-dominated sources could exist as unidentified extragalactic gamma-ray sources \citep[e.g.,][]{mas17}. On the other hand, highly synchrotron-dominated sources could exist as bright optical/radio blazars not detected by Fermi/LAT \citep[e.g.,][]{lis15}. We should also note that comparing single-band observed luminosities (W1 and 1~GeV) with the bolometric luminosities predicted by our model ignores the bolometric correction $L_{\rm syn}/L_{\rm W1}$, and the combination of bolometric corrections that affect Compton dominance $q_{\rm FW}$. Taking this into account, we can expect that the total jet powers predicted by our reference model could be underestimated by a factor of a few.

The results of our simplified analysis suggest that the emitting regions of blazar jets are out of equipartition, with $P_{\rm B}/P_{\rm e} \sim 0.2$. This conclusion is more solid in the case of BL Lacs, since $q_{\rm SSC}$ does not depend directly on the jet Lorentz factor or opening angle. It is encouraging that this conclusion is confirmed by results of SED modelling for BL Lacs \citep{tav16}. That this conclusion should apply equally to FSRQs is motivated mainly by our implicit assumption that the composition of all blazar jets should be comparable. In our picture, the transition between BL Lacs and FSRQs is solely due to systematic increase of the luminosity of external radiation fields with increasing accretion rate. Should our conclusion hold, it has strong implication for the dissipation physics of relativistic blazar jets: dissipation models based on magnetic reconnection in magnetically dominated jet regions \citep{gia09,nal11,pet16} would be disfavoured (at least for general, non-flaring emission), and instead models based on shock waves or turbulence in weakly magnetised relativistic plasma would merit a rethink.

\begin{acknowledgement}
We thank Marek Sikora for comments on the manuscript.
This work was supported by the Polish National Science Centre grant 2015/18/E/ST9/00580.
\end{acknowledgement}


\clearpage

\end{document}